# Deep ANN-based Touchless 3D Pad for Digit Recognition

Pramit Kumar Pal, Debarshi Dutta, Attreyee Mandal, Dipshika Das
Techno International New Town, Kolkata, India
pramitpalkumar@gmail.com, debarshidutta.official@gmail.com, attreyee24@gmail.com, dipshikasen@gmail.com

***Abstract*** - **The Covid-19 pandemic has changed the way humans interact with their environment. Common touch surfaces such as elevator switches and ATM switches are hazardous to touch as they are used by countless people every day, increasing the chance of getting infected. So, a need for touch-less interaction with machines arises. In this paper, we propose a method of recognizing the ten decimal digits (0-9) by writing the digits in the air near a sensing printed circuit board using a human hand. We captured the movement of the hand by a sensor based on projective capacitance and classified it into digits using an Artificial Neural Network. Our method does not use pictures, which significantly reduces the computational requirements and preserves users' privacy. Thus, the proposed method can be easily implemented in public places.**

*Index Terms* - artificial neural network, gesture recognition, microcontroller, projective capacitance.

## INTRODUCTION

The Covid-19 pandemic has put many restrictions on the daily lives of humans. Due to the highly transmissible nature of the virus, people need to sanitize their hands constantly. The chances of getting the virus from sources frequently used by the general public are very much possible. These sources are not limited to keypads in ATMs or keypads present in an elevator; a surface that people with their fingers frequently touch raises a need for a new touchless system for interacting with such systems.

We have developed a touchless system for entering numeric digits by use of mutual capacitance technology. Mutual capacitance is usually used for touch screens in mobile devices. However, by appropriate modification of the technology, we can use it for touchless human-computer interaction. The system we have developed is capable of registering 3D human hand gestures [10]. The use of an Artificial Neural Network makes it robust and reusable, i.e., the same system can be used for recognizing other symbols like numeric digits. This method can be an alternative to the conventional methods of inputting digits in a system.

## RELATED WORKS

The exploitation of capacitance property for human-computer interfacing is not new. It has been in use for decades. Today, almost all modern smartphones, tablet PCs, and computers with a touch screen use projected capacitance technology for the touch screen. Another everyday use of capacitance property for the human-computer interface is touchpads based on the capacitive sensing technique. In [3], Raphael et al. used capacitive sensing methods for performing simple gestures like picking or navigating objects on a screen. The authors placed four electrodes around the four sides of a tablet screen. In [2], Lee et al. built a dodecahedron-shaped input device that provided 24 degrees of freedom. Each of the 12 flat surfaces consisted of a capacitive sensor. The authors demonstrated the devices' feasibility in three-dimensional modeling tasks.

In [1], the authors made use of electric field sensing to track human hands. In [15], the author used inertial sensors in smartphones to interpret characters by performing the characters' mid-air gestures, which lacked visual feedback to the user performing the gestures mid-air. Reference [7], however, used motion tracking and a hidden Markov model for interpreting characters performed by the user. In [9], the author used cameras for recognition of writing characters in mid-air and reference. The downside of using cameras is that they cannot be used in extreme lighting conditions (either very dark or bright). Thus, they require optimum lighting conditions for proper functioning. Camera sensors are expensive to replace as their manufacturing costs are high. The use of cameras also poses a risk to the user's privacy. In references [8] and [6], the user wore a device on their wrist to recognize gestures performed by them.

In [14], the authors studied extensive area array sensing feasibility using projected capacitance as the underlying principle. In [5], the authors developed an 8X8 capacitive sensor array to perform gestures for rehabilitation purposes. Here, the capacitive sensor array uses mutual capacitance as the underlying technology.

The previous works primarily focus on detecting hand gestures from a visual data feed using camera-based gesture recognition systems and capacitive sensing to either track hand movements or recognize predefined hand gestures. Some works focused on getting gesture input from the user with devices worn on their hand by the user. As has been done in this paper, the projected capacitance technique for recognizing alphanumeric characters is, to our knowledge, completely new.

The recognized characters can then be sent to another machine where alphanumeric inputs are necessary for operation.

## BACKGROUND

Projected capacitance [5] technology is regularly used in modern touch screen devices like smartphones and tablets. However, in these devices, the sensing range is minimal, and thus they require physical touches to initiate an action. Since touch screens only detect finger positions when they are touched, by suitably modifying the technique and using deep learning [13], projected capacitance can be used to recognize touch-less hand gestures [6], which is the main contribution of this paper. Fig: I show the flow diagram for our system from getting gesture input from the user and classifying the gestures into digits [14].

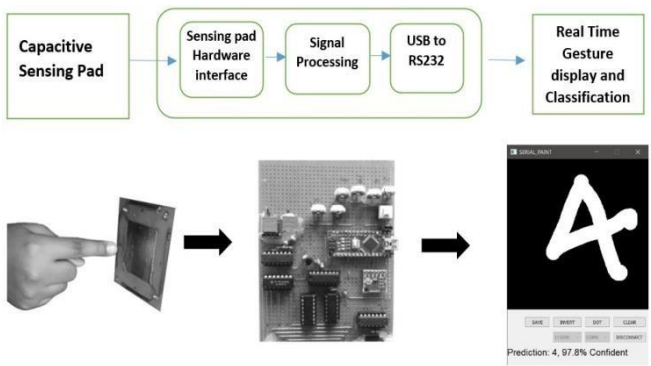

FIGURE I
FLOW DIAGRAM OF WORKING PROTOTYPE

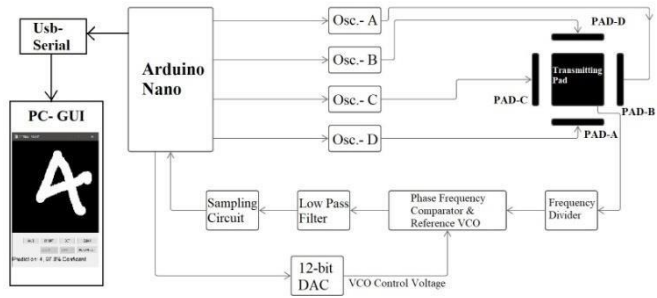

FIGURE II
BLOCK DIAGRAM OF HARDWARE PROTOTYPE

## PROTOTYPE OVERVIEW

Our prototype consists of three main parts: sensing printed circuit board, Processing board, and Communication with PC via USB to RS232 converter. The complete block diagram representation of the prototype is shown in Fig: II.

The sensing printed circuit board consists of 5 electrodes etched on a copper-clad board made of the FR4

substrate material. This printed circuit board is responsible for sensing the position of the user's hand in front of it. As the etched electrodes form mutual capacitance, it is susceptible to small changes in capacitance due to the presence of a human hand near it [3]. The processing board processes this small change in capacitance and then fed to the Arduino Nano to communicate with the PC interface via USB to RS232 converter at a baud rate of 115200. The mutual capacitor formed by the sensing electrodes is part of an oscillator whose frequency is influenced by the presence of the human hand when it is within the range of sensing of the electric fields emanating from the electrodes [1]. When inside the sensing region, the hand cuts out electrostatic field lines and diverts them to another direction so that the electric flux density falling on the receiving electrodes decreases, as shown in Fig: III

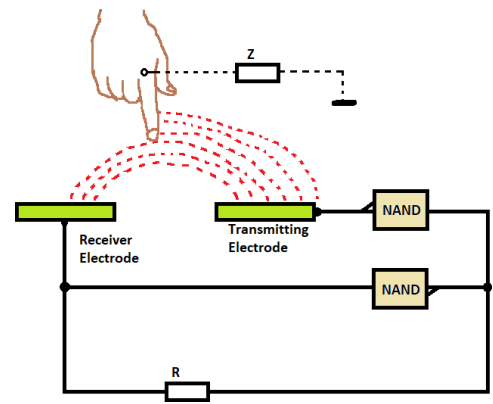

FIGURE III
DIVERSION OF ELECTRIC FIELD LINES BY THE HAND

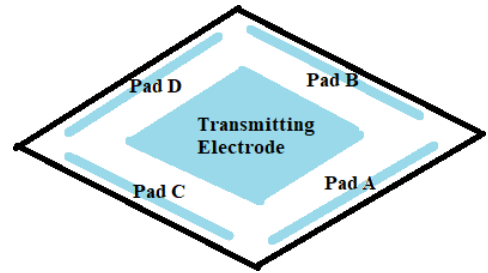

FIGURE IV
SENSING PRINTED CIRCUIT BOARD WITH FIVE ELECTRODES

From the five electrodes etched on the sensing printed circuit board, shown in Fig: IV, four electrodes surrounding the center electrode are connected to four separate oscillators' high impedance input. These four electrodes are called receiving electrodes named Pad-A, Pad-B, Pad-C, and Pad-D. They are responsible for giving spatial information about the position of the human hand. The center electrode or the transmitting electrode is connected to a low impedance output of the oscillators by a NAND gate.

When the hand gradually approaches the electrodes, the capacitance between the electrodes decreases, increasing the frequency of the oscillators. As we need three-dimensional coordinates from the system, each of the five electrodes must be scanned one after another in each sampling cycle. One such sampling cycle is represented in Fig: V.

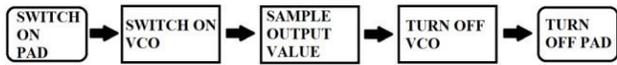

FIGURE V
ONE SAMPLING CYCLE FOR A SINGLE SENSING PAD

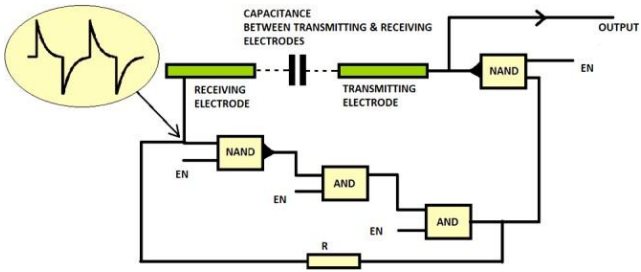

FIGURE VI
BLOCK DIAGRAM OF AN OSCILLATOR FOR A SINGLE ELECTRODE

The five electrodes are scanned 80 times per second or at a sample rate of 80Hz. Fig: VI show one such oscillator for a single electrode.

The oscillators are turned on by logic high at the ENABLE pin of the oscillator shown in Fig: VI. Each oscillator is tuned to a frequency of 1.7 MHz by the resistor R shown in Fig: VI. Under the influence of the human hand, the oscillators' frequency changes by a few hundred parts per million. (100 parts per million =0.01%). To track the position of the hand, this small change in frequency needs to be measured accurately, is achieved by a phase-frequency comparator, which compares the frequency of the four oscillators with a reference oscillator, realized by a voltage control oscillator and converts this change in frequency to a voltage signal. Unlike small changes in capacitance, the voltage changes can be measured easily by the ADC module of the Arduino Nano [12].

The oscillators' output is scaled down by a frequency divider, as shown in Fig: II so that the reference oscillator and the sensing oscillators are in the same frequency range. Due to the susceptible nature of the oscillators, they are susceptible to environmental factors like input voltage and temperature. The reference frequency needs to be adjusted by small amounts so that the output from the phase-frequency comparator is minimal when a human hand is not present. The Arduino Nano adjusts the reference frequency individually for each pad before scanning by sending a control voltage to the voltage-controlled oscillator by a 12-bit DAC. The output of the phase-frequency comparator has a high-frequency component at its output, filtered out by a simple first-order RC low pass filter with a cutoff frequency of 72.3Hz to capture very tiny movements of the hand yet filter out unwanted high-frequency noise.

During scanning or multiplexing, each of the four receiving electrodes the sample and hold circuit samples and holds the filtered output voltage of the phase-frequency comparator, which is then read by the analog to digital converter of the Arduino Nano. All the processes, from setting the frequency of the reference oscillator, scanning each electrode, to communicating with the graphical user interface running in the personal computer, are done by the Arduino Nano. The Arduino Nano consists of an 8-bit ATMEGA328 microcontroller running at 16MHz.

## GESTURE CAPTURE

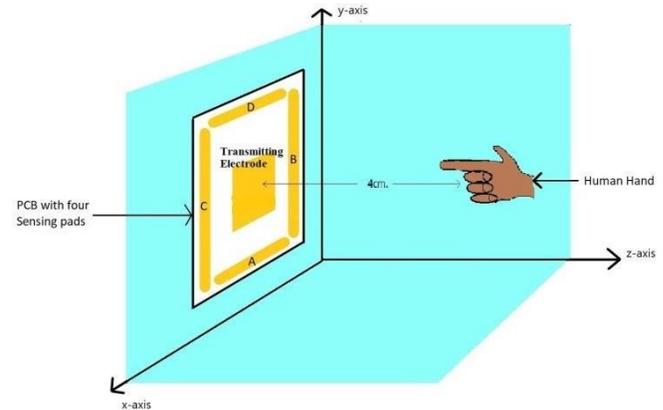

FIGURE VII
HAND NEAR SENSING PRINTED CIRCUIT BOARD

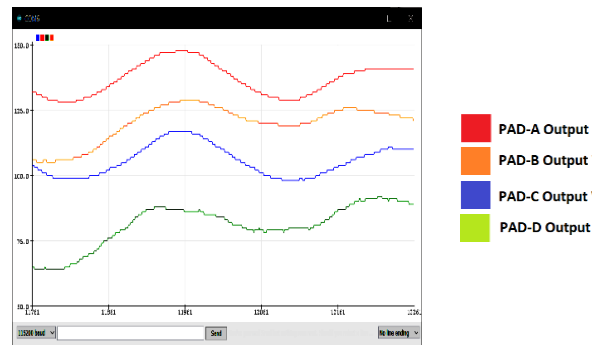

FIGURE VIII
OUTPUT FROM FOUR RECEIVING ELECTRODES

We can get three-dimensional coordinates for the static position of the hand from the outputs of these four sensing pads, as shown in Fig: VII, where a human hand is in the sensing range of the pads. The output of each pad, namely Pad A, Pad-B, Pad-C, and Pad-D, can be seen in Fig: VIII as output values that can be differentiated from each other by their colors. From the three-dimensional coordinates, the Graphical User Interface shown in Fig: X can draw the gestures performed by the user in front of the sensing pads in real-time as the hand is moved in three-dimensional space by the user as depicted in Fig: VII.

The graphical user interface [1] for visualization of the gestures drawn by the user in real-time was developed using Python 3.9 and PyQt5 libraries. The algorithm for the GUI interface is shown in Fig: IX.

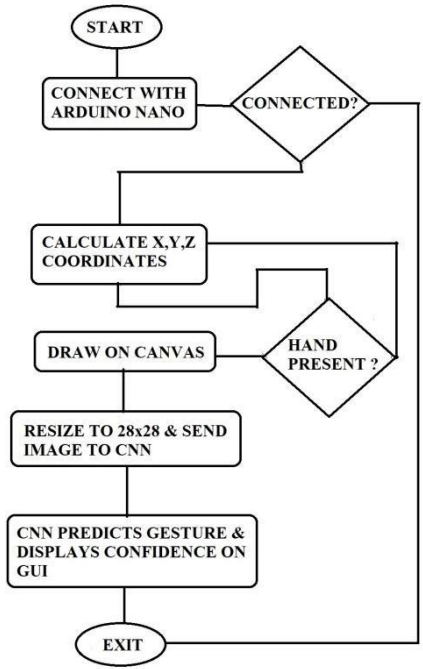

FIGURE IX
FLOWCHART FOR THE GRAPHICAL USER INTERFACE

The calculation of the X Y and Z coordinates from the raw output signal of the four electrodes are given as follows:

$$X \ Coordinate = (A - D) \tag{1}$$

$$Y \ Coordinate = (B - C) \tag{2}$$

$$Z \ Coordinate = (A + B + C + D)/4 \tag{3}$$

*Where A, B, C, D represents the output of each of four electrodes.*

The X, Y, and Z coordinates are calculated by the equation given in equations (1), (2), and (3), respectively.

The graphical user interface starts drawing the gestures on the interface canvas when the user has brought his/her hand at a distance of 4 cm or less from the sensing pads. The gesture is recognized as complete when the user's hand is more than 4 cm from the sensing electrodes. Upon completion of each gesture, the software automatically captures a screenshot of the canvas region of the graphical user interface, which can be seen in Fig: X as the black region. The captured image is converted to grayscale color format and resized to an image size of (28 x 28x1) pixels.

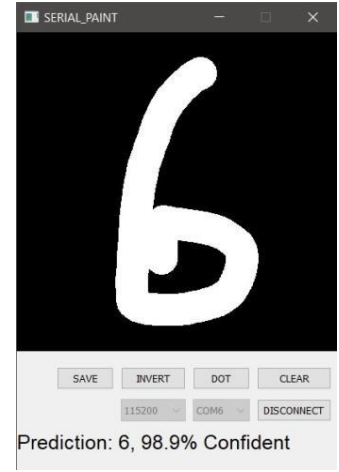

FIGURE X
A SCREENSHOT OF THE GRAPHICAL USER INTERFACE FOR OUR PROTOTYPE

## TRAINING AND TESTING DATASET

We have created our dataset for training and testing various models and comparing them with the best performance. The 50000 images for the digit classification dataset contain ten classes (0–9), each of which is 28x28 pixels in size. Each class comprises 5000 grayscale images. The dataset has been divided into training and testing in the ratio of 8:2, respectively.

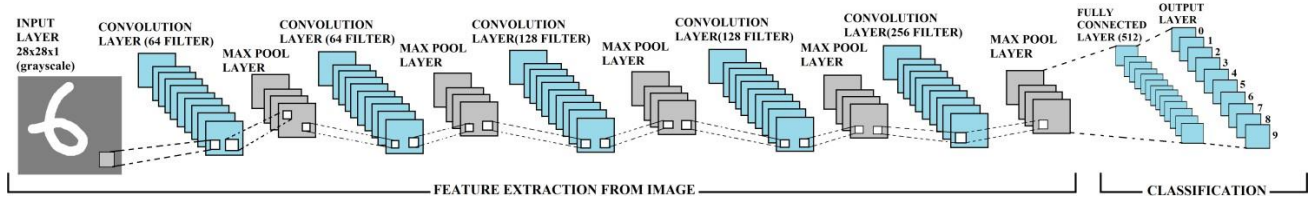

FIGURE XI
STRUCTURE OF CNN MODEL

## DEEP LEARNING MODELS

### DEEP CNN WITH DATA AUGMENTATION

We propose a deep learning model for classifying gestures as digits (0-9). This model is developed based on CNN (Convolutional Neural Network) architecture [11]. CNN is a deep learning model with widespread use in computer vision and image classification. Fig: XI shows a diagrammatic representation of the various layers of the model.

Our model has an input layer of 28x28x1 followed by multiple sub-layers of filters, max-pooling layers, batch normalization layers, and a densely connected network.
   RELU (Rectified Linear Unit) layer was chosen as an activation function after each convolution layer due to its nonlinear nature.

$$f(x) = \max(0, x) \qquad (4)$$

The equation for the RELU activation function is given in equation (4). The dense output layer uses the SoftMax activation function, shown in equation (5), to obtain the probability of different classes.

$$P(y = j|\theta^i) = \frac{e^{\theta^i}}{\sum\limits_{j=0}^{k} e^{\theta^i_k}} \qquad (5)$$

Where $\theta = w_0 x_0 + w_1 x_1 + \cdots + w_k x_k = \sum_{i=1}^{k} w_i x_i = w^T x$

The CNN model is shown in Fig: XI is trained on the augmented dataset by Adam optimizer with a learning rate of 0.001 and mini-batch size of 64. The augmentation involves random rotation of images within 0-degree to 180-degree, random zooming in on the images, and shifting the height and width of the images. Data augmentation ultimately introduced more variation in the dataset and thus helped the model to perform better. Fig: XIV shows the epoch versus loss and accuracy of the CNN model, and Fig: XVIII shows the confusion matrix for the validation data.

### DEEP CNN WITHOUT DATA AUGMENTATION

The deep CNN model proposed in Fig: XI was trained and validated with our training and testing data, but withoutdata-augmentation applied to the dataset. Fig: XV shows the epoch versus loss and accuracy of the CNN model without data augmentation, and Fig: XIX shows the confusion matrix for the validation data. The sequential layers in this CNN model were the same as the model proposed above in Fig: XI.

## MULTI-LAYER PERCEPTRON

Multi-layer Perceptron (MLP) is a type of artificial neural network widely used for classification and regression tasks. The task of classifying images is complex and requires complex deep learning models. The pictorial representation of the MLP model evaluated by us is shown in Fig: XII. MLP is trained using Adam optimizer, with a learning rate of 0.001 and batch size 32.

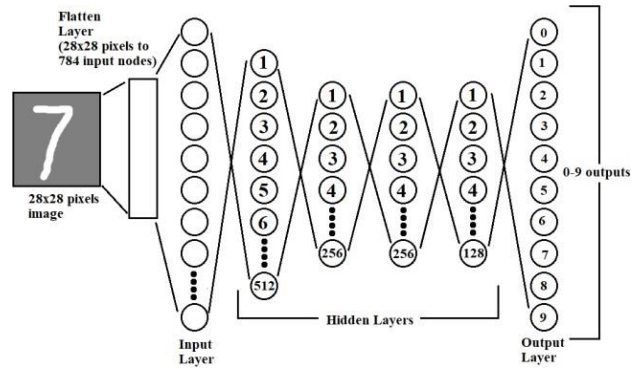

FIGURE XII
MULTI-LAYER PERCEPTRON MODEL LAYERS

### RNN WITH DATA AUGMENTATION

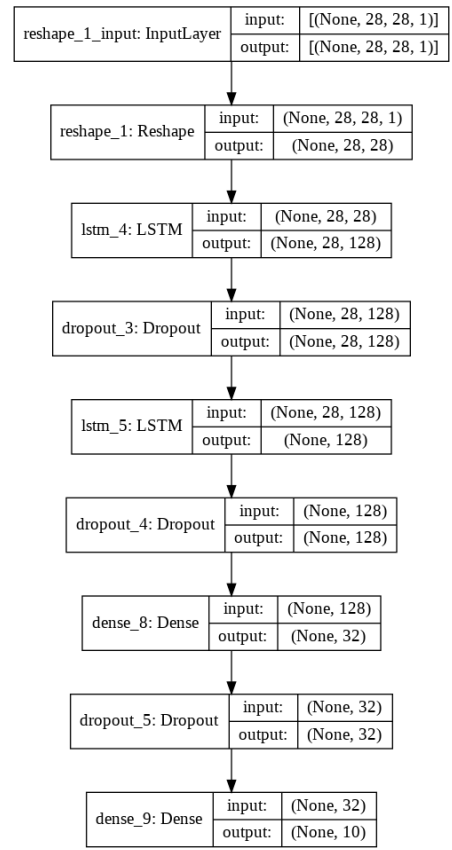

FIGURE XIII
RNN WITH MODEL SUMMARY

RNN [11] is a deep learning model with memory mainly used to process time-series data. Here we have used unidirectional LSTM (Long Short-Term Memory), which is a type of RNN. The layers of the RNN model we evaluated are shown in Fig: XIII.

The training was done by Adam optimizer, with a learning rate of 0.001 and batch size of 32.

The data augmentation involves random rotation of images within 0-degree to 180-degree, random zooming in on the images, and shifting the height and width of the images.

## RESULTS AND ANALYSIS

In an experiment, we performed training and testing on different deep learning models and compared those results. The different architectures on which the comparisons are made are as follows:

- CNN with Data Augmentation
- CNN without Data Augmentation
- RNN
- MLP (Multi-Layer Perceptron)

The training and testing performance for the CNN model with data augmentation is shown in Fig: XIV. Fig: XV shows the training and testing performance for the CNN model without data augmentation. Fig: XVI shows the training and testing performance for RNN with data augmentation applied, and Fig XVII shows the training and testing performance of the MLP model

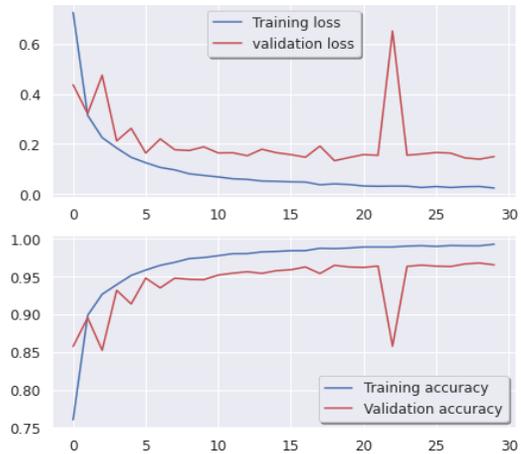

FIGURE XV
TRAINING AND TESTING PERFORMANCES FOR CNN WITHOUT DATA AUGMENTATION

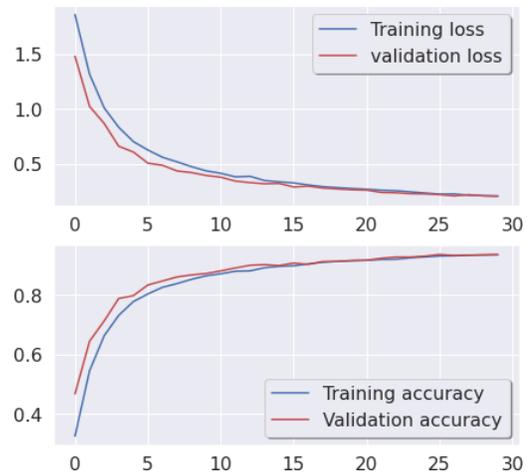

FIGURE XVI
TRAINING AND TESTING PERFORMANCES FOR RNN WITH DATA AUGMENTATION

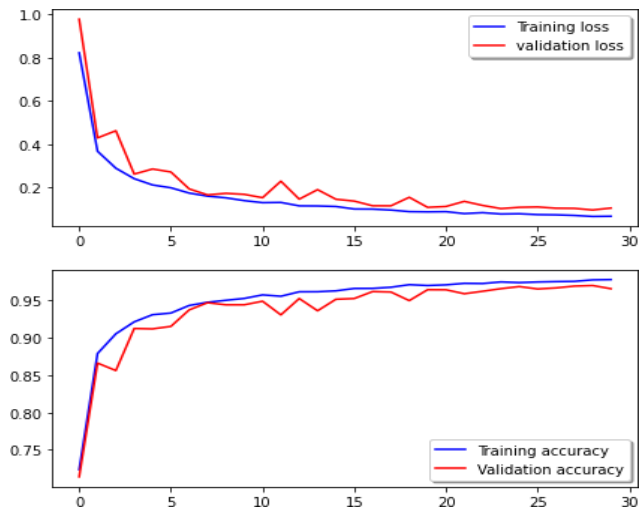

FIGURE XIV
TRAINING AND TESTING PERFORMANCES OF CNN WITH DATA AUGMENTATION

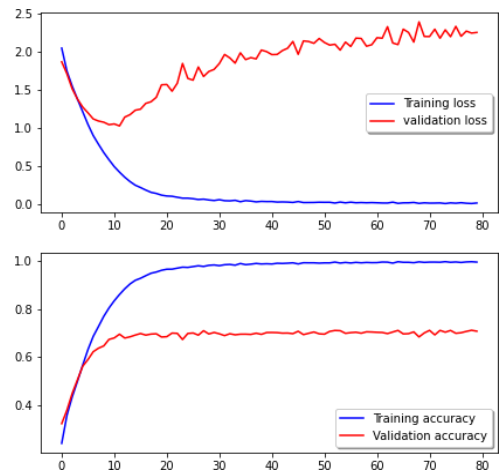

FIGURE XVII
TRAINING AND TESTING PERFORMANCES FOR MLP

The confusion matrices for the different models are shown in Fig: XVIII for CNN with data augmentation, Fig: XIX for CNN model without data augmentation, Fig: XX for RNN with data augmentation, and Fig: XXI for the MLP model. The density of colors in each grid represents the number of correct or incorrect representations for the corresponding models.

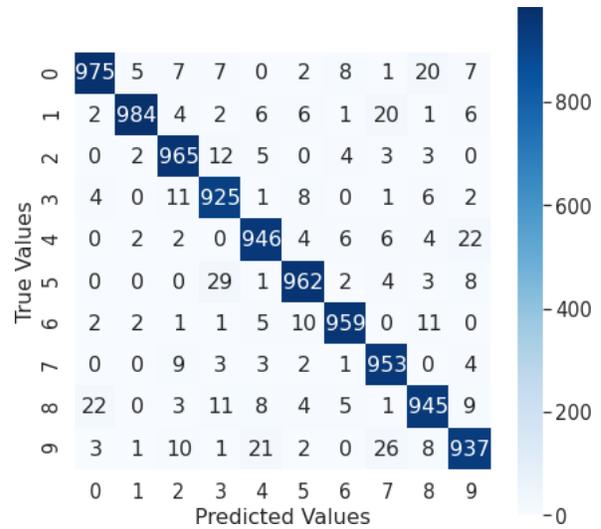

**FIGURE XX**
CONFUSION MATRIX FOR RNN WITH DATA AUGMENTATION

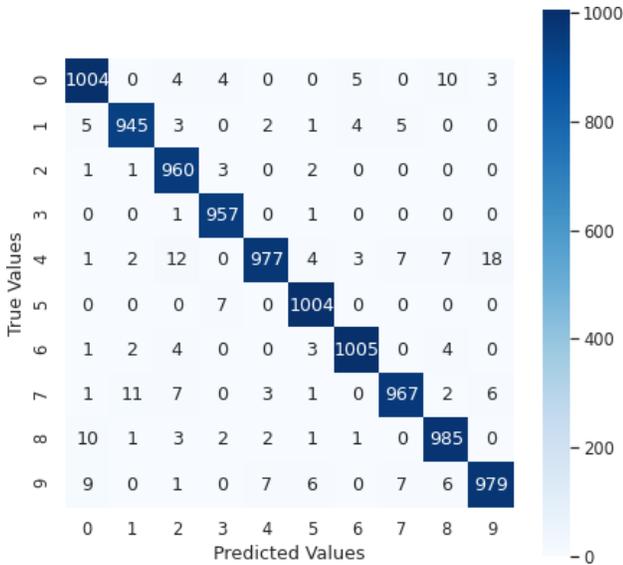

**FIGURE XVIII**
CONFUSION MATRIX FOR CNN WITH DATA AUGMENTATION

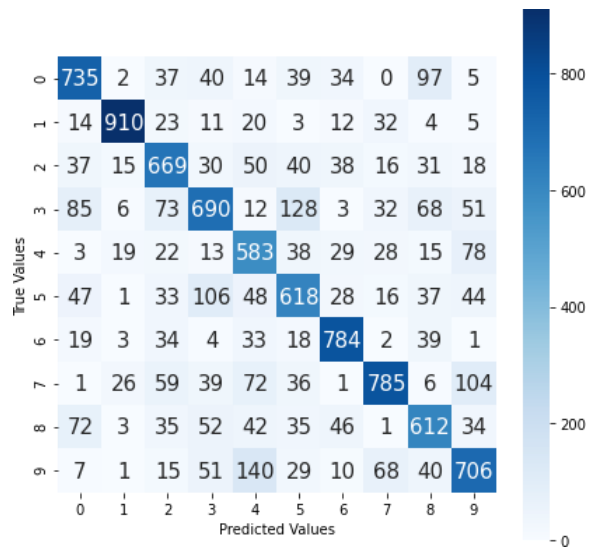

**FIGURE XXI**
CONFUSION MATRIX FOR MLP

## SUMMARY OF DIFFERENT ARCHITECTURE

**TABLE I**
ACCURACY AND LOSS FOR DIFFERENT MODELS

| Model | Optimizer | Train Loss | Train Acc. | Validation Loss | Validation Acc. |
|---|---|---|---|---|---|
| CNN with data Augmentation | Adam | 0.0641 | 0.9782 | 0.0967 | 0.9703 |
| CNN without data augmentation | Adam | 0.0223 | 0.9936 | 0.1497 | 0.9656 |
| RNN | Adam | 0.3125 | 0.9050 | 0.3042 | 0.9023 |
| MLP | Adam | 0.0171 | 0.9950 | 2.2557 | 0.7080 |

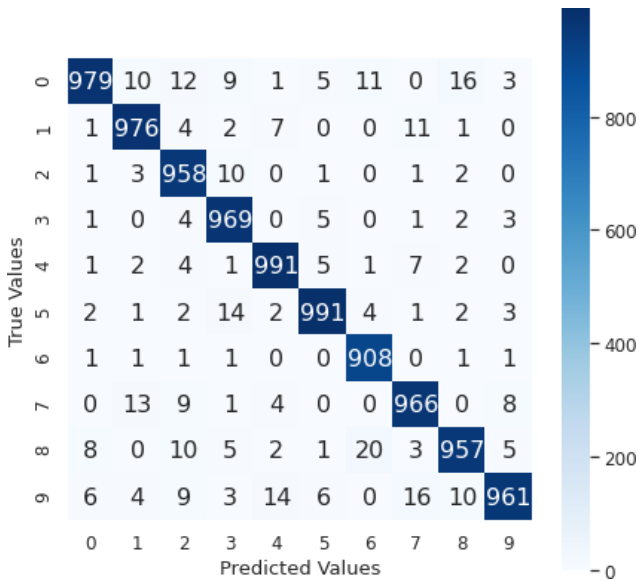

**FIGURE XIX**
CONFUSION MATRIX FOR CNN WITHOUT DATA AUGMENTATION

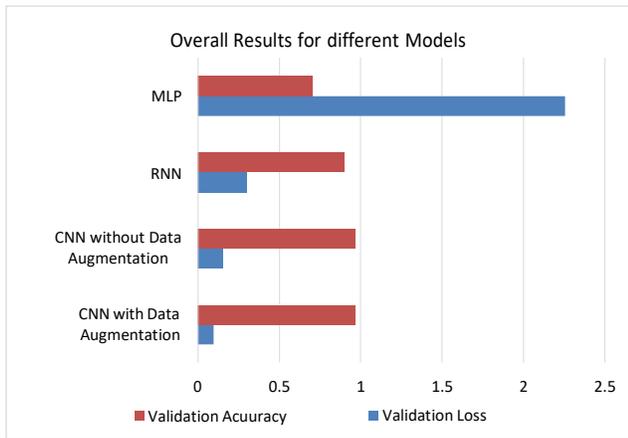

**FIGURE XXII**
CLUSTERED BAR CHART FOR VALIDATION ACCURACY AND VALIDATION LOSS OF DIFFERENT MODELS

Fig: XXII shows the comparison between different models in the form of a clustered bar chart. Table 1 shows the comparison in performance between different models.

From the comparison between different models, the CNN model with data augmentation performed best in both aspects of classification accuracy and validation loss, having 97.03% and 0.0967, respectively, among other models, with or without data augmentation. For the same CNN model, if data augmentation was not applied, the model suffered from fitting, which is observable from Fig: XV.

Thus, for the final prototype, the CNN model with data augmentation was selected for best performance. The trained model was exported and used in live classification software based on a graphical user interface, receiving data from the hardware prototype and drawing inference on which the user drew digits.

The confidence percentage of each classification result shows how accurately the model recognizes the digits based on the user's gestures. Such an instance of digit classification is shown in Fig: X.

## CONCLUSION

This paper presents a novel system to enable touch-less hand gesture recognition for digit classification using a deep learning model. Different ANNs with data augmentation were proposed, which classified digits based on user gestures. The comparisons between various models show that CNN with data augmentation performs best in classification accuracy and reliability. Future work can involve tracking and recognition of more complex gestures like alphabets and special characters. Optical RGB sensors have been widely used for gesture recognition. The fusion of multiple sensors could thus obtain more accuracy and reliability, thus increasing system performance. Our approach can be used in systems where touching a surface is unhygienic and dangerous, considering the current ongoing global COVID-19 pandemic.